\documentclass[aps,prd,twocolumn,floatfix,superscriptaddress,balancelastpage,nofootinbib,amsmath,showpacs]{revtex4-1}
\usepackage{graphicx,epstopdf,epsfig,bm,dcolumn,natbib,amsmath,amssymb}

\def\lsim{\mathrel{\raise.3ex\hbox{$<$\kern-.75em\lower1ex\hbox{$\sim$}}}}
\def\gsim{\mathrel{\raise.3ex\hbox{$>$\kern-.75em\lower1ex\hbox{$\sim$}}}}

\begin{document}

\title{Hidden Sector Dark Matter Models for the Galactic Center Gamma-Ray Excess}

\author{Asher Berlin}
\affiliation{Department of Physics, University of Chicago, Chicago, IL 60637}
\author{Pierre Gratia}
\affiliation{Department of Physics, University of Chicago, Chicago, IL 60637}
\author{Dan Hooper}
\affiliation{Center for Particle Astrophysics, Fermi National Accelerator Laboratory, Batavia, IL 60510}
\affiliation{Department of Astronomy and Astrophysics, University of Chicago, Chicago, IL 60637}
\author{Samuel D. McDermott}
\affiliation{Center for Particle Astrophysics, Fermi National Accelerator Laboratory, Batavia, IL 60510}
\affiliation{Michigan Center for Theoretical Physics, Ann Arbor, MI 48109}

\begin{abstract}

The gamma-ray excess observed from the Galactic Center can be interpreted as dark matter particles annihilating into Standard Model fermions with a cross section near that expected for a thermal relic. Although many particle physics models have been shown to be able to account for this signal, the fact that this particle has not yet been observed in direct detection experiments somewhat restricts the nature of its interactions. One way to suppress the dark matter's elastic scattering cross section with nuclei is to consider models in which the dark matter is part of a hidden sector. In such models, the dark matter can annihilate into other  hidden sector particles, which then decay into Standard Model fermions through a small degree of mixing with the photon, $Z$, or Higgs bosons. After discussing the gamma-ray signal from hidden sector dark matter in general terms, we consider two concrete realizations: a hidden photon model in which the dark matter annihilates into a pair of vector gauge bosons that decay through kinetic mixing with the photon, and a scenario within the generalized NMSSM in which the dark matter is a singlino-like neutralino that annihilates into a pair of singlet Higgs bosons, which decay through their mixing with the Higgs bosons of the MSSM. 

\end{abstract}

\pacs{95.85.Pw, 98.70.Rz, 95.35.+d; FERMILAB-PUB-14-134-A, MCTP-14-12} 

\maketitle

\section{Introduction}

Data from the \textit{Fermi} Gamma-Ray Space Telescope has been found to contain a highly statistically significant signal from the region surrounding the Galactic Center, with a spectrum and angular distribution compatible with that anticipated from annihilating dark matter particles~\cite{Goodenough:2009gk,Hooper:2010mq,Boyarsky:2010dr,Hooper:2011ti,Abazajian:2012pn,Gordon:2013vta,Hooper:2013rwa,Huang:2013pda,Abazajian:2014fta,Daylan:2014rsa}. In particular, the recent analysis of Ref.~\cite{Daylan:2014rsa} found the spectrum of this signal to be well-fit by 31-40 GeV dark matter particles annihilating to $b$-quarks, or somewhat lower mass dark matter particles annihilating to $c\bar{c}$, $s\bar{s}$, $d\bar{d}$ or $u\bar{u}$.  The morphology of the signal is spherically symmetric with respect to the Galactic Center, and falls off at a rate that is consistent with a dark matter halo profile described by $\rho \propto r^{-\gamma}$, with $\gamma\simeq 1.1-1.3$. The signal is not confined to the central stellar cluster, but can be identified out to angles exceeding $10^{\circ}$ from the Galactic Center. Furthermore, the annihilation cross section required to normalize the observed signal is $\sigma v \sim 2\times 10^{-26}$ cm$^3$/s, in good agreement with that predicted for dark matter in the form of a simple thermal relic. And although astrophysical explanations for this signal have been proposed (including a large population of unresolved millisecond pulsars~\cite{Hooper:2010mq,Abazajian:2010zy,Hooper:2011ti,Abazajian:2012pn,Gordon:2013vta,Abazajian:2014fta} or cosmic-ray interactions with gas~\cite{Hooper:2010mq,Hooper:2011ti,Abazajian:2012pn,Gordon:2013vta}), none of these proposals appear to be viable in light of the most recent observations~\cite{Linden:2012iv,Hooper:2013rwa,Huang:2013pda,Hooper:2013nhl,Macias:2013vya}. 

Several groups have studied dark matter models potentially responsible for the gamma-ray excess~\cite{Berlin:2014tja,Izaguirre:2014vva,Alves:2014yha,Agrawal:2014una,Cerdeno:2014cda,Ipek:2014gua,Ghosh:2014pwa,Ko:2014gha,Boehm:2014bia,Abdullah:2014lla,Martin:2014sxa} (for earlier work, see Refs.~\cite{Boehm:2014hva,Modak:2013jya,Huang:2013apa,Okada:2013bna,Hagiwara:2013qya,Buckley:2013sca,Anchordoqui:2013pta,Buckley:2011mm,Boucenna:2011hy,Marshall:2011mm,Zhu:2011dz,Buckley:2010ve,Logan:2010nw}), and many scenarios have been identified in which dark matter annihilating directly to Standard Model fermions can generate a gamma-ray signal with the observed characteristics.  Despite the fact that the null results of direct detection experiments significantly restrict the nature of the dark matter's interactions with quarks, a sizable fraction of models capable of accommodating the gamma-ray excess predict elastic scattering cross sections with nuclei that are compatible with these constraints~\cite{Berlin:2014tja,Izaguirre:2014vva,Alves:2014yha}.

Alternatively, one might take the lack of signals in direct detection experiments as motivation to consider models in which the dark matter is not charged under the Standard Model gauge group, but instead is part of a hidden sector~\cite{Ko:2014gha,Boehm:2014bia,Abdullah:2014lla,Martin:2014sxa}. In particular, one could consider models in which the dark matter annihilates into particles that couple to the Standard Model only through a small degree of (mass or kinetic) mixing.
Such models give rise to $2 \to 3$ or $2 \to 4$ annihilation diagrams and can have rather different phenomenology than the $2\to 2$ models described above.  In particular, interactions between dark matter and nuclei can be highly suppressed, and the prospects for studying such models at the LHC are in general less encouraging than in models with direct couplings to the Standard Model.

\begin{figure*}[t]
\begin{center}
\includegraphics[width=0.6\textwidth]{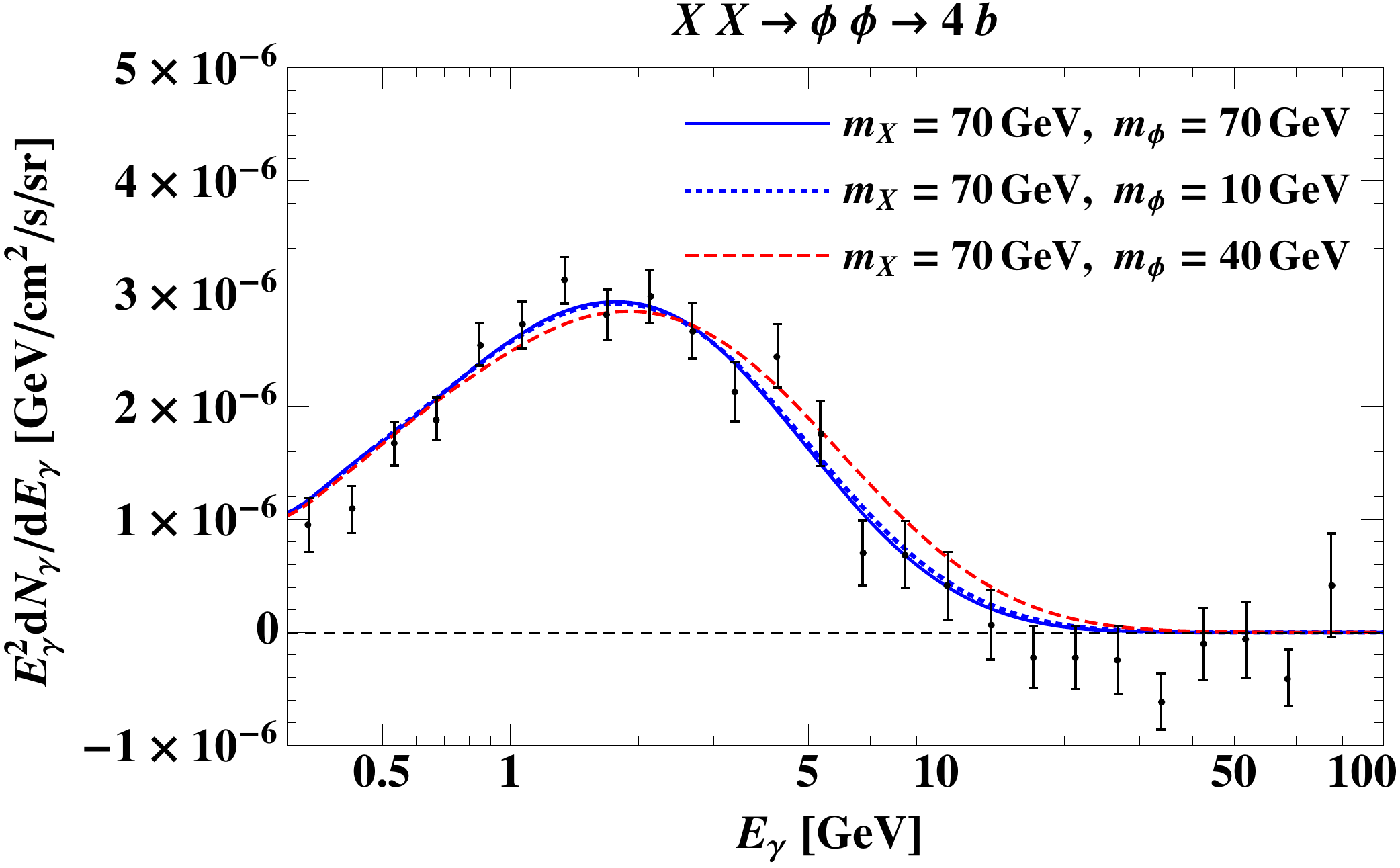}
\caption{The shape of the gamma-ray spectrum produced by the annihilations of 70 GeV dark matter particles into a pair of 70, 40 or 10 GeV intermediate states, which then each decay into $b\bar{b}$. This is compared to the spectrum of the observed gamma-ray excess, as reported in Ref.~\cite{Daylan:2014rsa}.}
\label{bbspectra}
\end{center}
\end{figure*}

In this paper, we consider the possibility that the Galactic Center's gamma-ray excess originates from dark matter particles that are part of a hidden sector. In Sec.~\ref{fit}, we calculate the spectrum of gamma-rays produced through dark matter annihilating into other hidden sector states that decay into Standard Model particles, and compare this to the spectrum observed from the Galactic Center. In Secs.~\ref{hiddenphoton} and~\ref{nmssm}, we consider two specific models which represent possible realizations of this phenomenology. In particular, we consider a model in which the dark matter annihilates into a pair of hidden photons (the massive vector bosons associated with a new broken $U(1)$ gauge group). We also consider a supersymmetric model with a hidden sector that consists of a complex Higgs singlet (corresponding to a physical scalar and pseudoscalar) and its superpartner (the singlino).  In Sec.~\ref{conclusion} we summarize our results and conclusions.

\section{Fitting the observed gamma-ray spectrum with cascade annihilations}
\label{fit}

Consider a pair of dark matter particles, $X$, annihilating at rest into two on-shell particles, $\phi_1$ and $\phi_2$, with masses $m_{\phi_1}$ and $m_{\phi_2}$, respectively. Energy and momentum conservation require the Lorentz factors of $\phi_1$ and $\phi_2$ to be given by:
\begin{align}
\begin{split}
\gamma_{1,2} &= \frac{s + m^2_{\phi_1,\phi_2}-m^2_{\phi_2,\phi_1}}{2 m_{\phi_1,\phi_2} \sqrt s}, \\
&\simeq \frac{4 m^2_X + m^2_{\phi_1,\phi_2}-m^2_{\phi_2,\phi_1}}{4 m_{\phi_1,\phi_2} \, m_X}.
\end{split}
\end{align}
In what follows, we will be interested in the case in which each $\phi_1$ and $\phi_2$ decay into Standard Model particles, producing a spectrum of gamma-rays in the $\phi$'s rest frame that we denote by $(dN_{\gamma}/dE_{\gamma})_{\phi_i}$. After boosting into the lab frame, each dark matter annihilation produces a gamma-ray spectrum given by:
\begin{equation}
\frac{dN_{\gamma}}{dE_{\gamma}} = \sum_{i=1,2}\frac{1}{2\beta_i \gamma_i} \int^{E_{\gamma}/\gamma_i(1-\beta_i)}_{E_{\gamma}/\gamma_i(1+\beta_i)}  \frac{dE'_{\gamma}}{E'_{\gamma}} \, \bigg(\frac{dN_{\gamma}}{dE'_{\gamma}}\bigg)_{\phi_i},
\end{equation}
where $\beta_i=(1-\gamma_i^{-2})^{1/2}$. We use {\tt PPPC4DMID} \cite{Cirelli:2010xx} for the photon spectrum from heavy quarks, light quarks, and leptons, and we use {\tt DarkSUSY} \cite{Gondolo:2004sc} for annihilation through $c \bar c$ states.

In Fig.~\ref{bbspectra}, we show the shape of the gamma-ray spectrum that results from the annihilation of 70 GeV dark matter particles into a pair of intermediate states with a common mass ($m_{\phi_1}=m_{\phi_2}$), each of which then decays into $b\bar{b}$. When this is compared to the spectrum of the gamma-ray excess observed from the Galactic Center (represented by error bars)~\cite{Daylan:2014rsa}, it is evident that a good fit can be obtained. Whereas direct annihilation to $b\bar{b}$ requires dark matter masses of 31-40 GeV to fit the observed spectrum~\cite{Daylan:2014rsa}, annihilations through an intermediate state favor dark matter masses which are roughly twice as large, with the precise value depending on the masses of the intermediate particles. 
In either scenario, the spectrum of the Galactic Center gamma-ray excess is well fit by $b \bar b$ pairs with Lorentz boosts of $\gamma \sim 7$.

\begin{figure*}[tb]
\begin{center}
\includegraphics[width=0.4\textwidth]{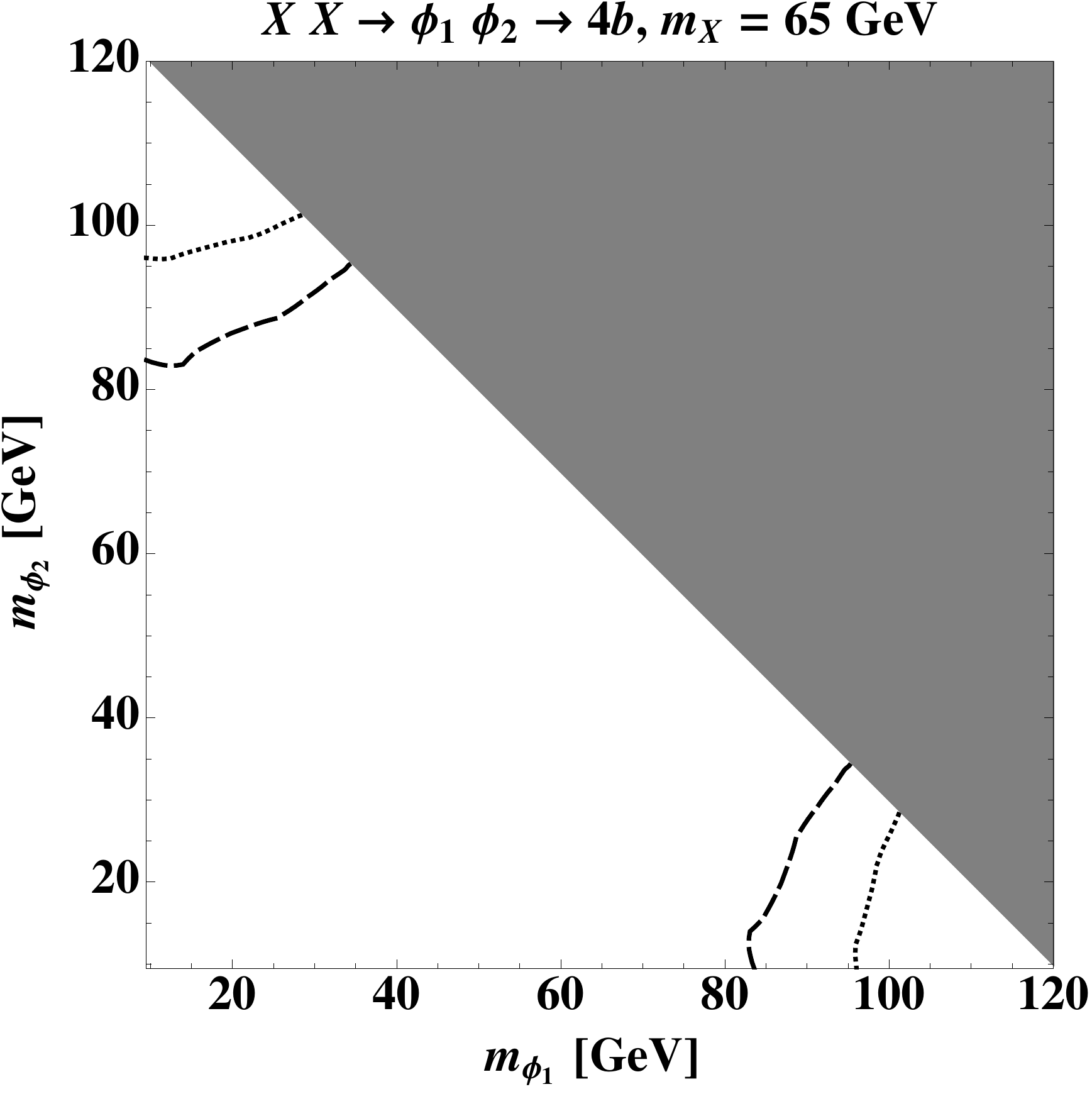}
\vspace{0.2cm}
\includegraphics[width=0.385\textwidth]{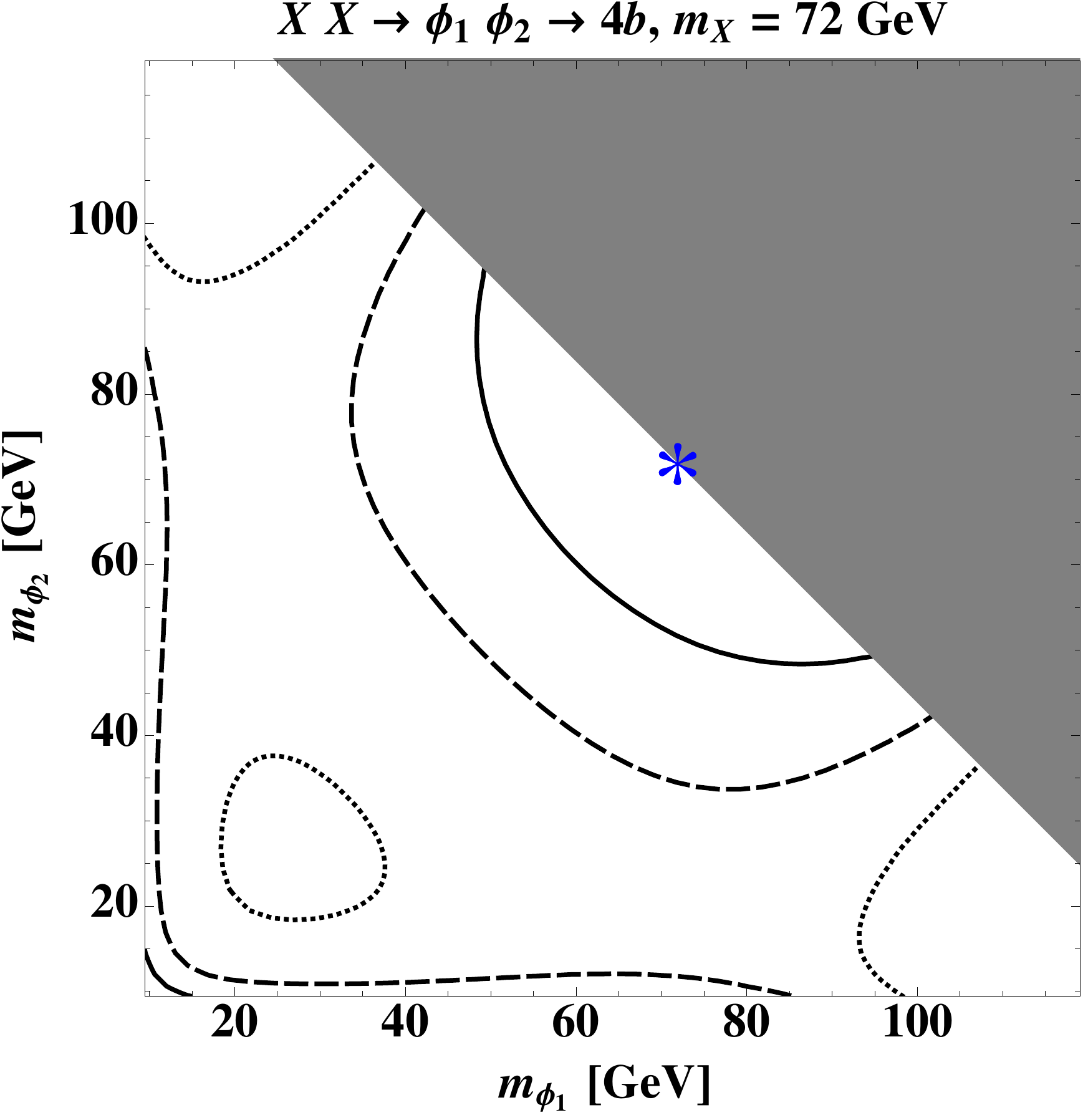}\\
\includegraphics[width=0.4\textwidth]{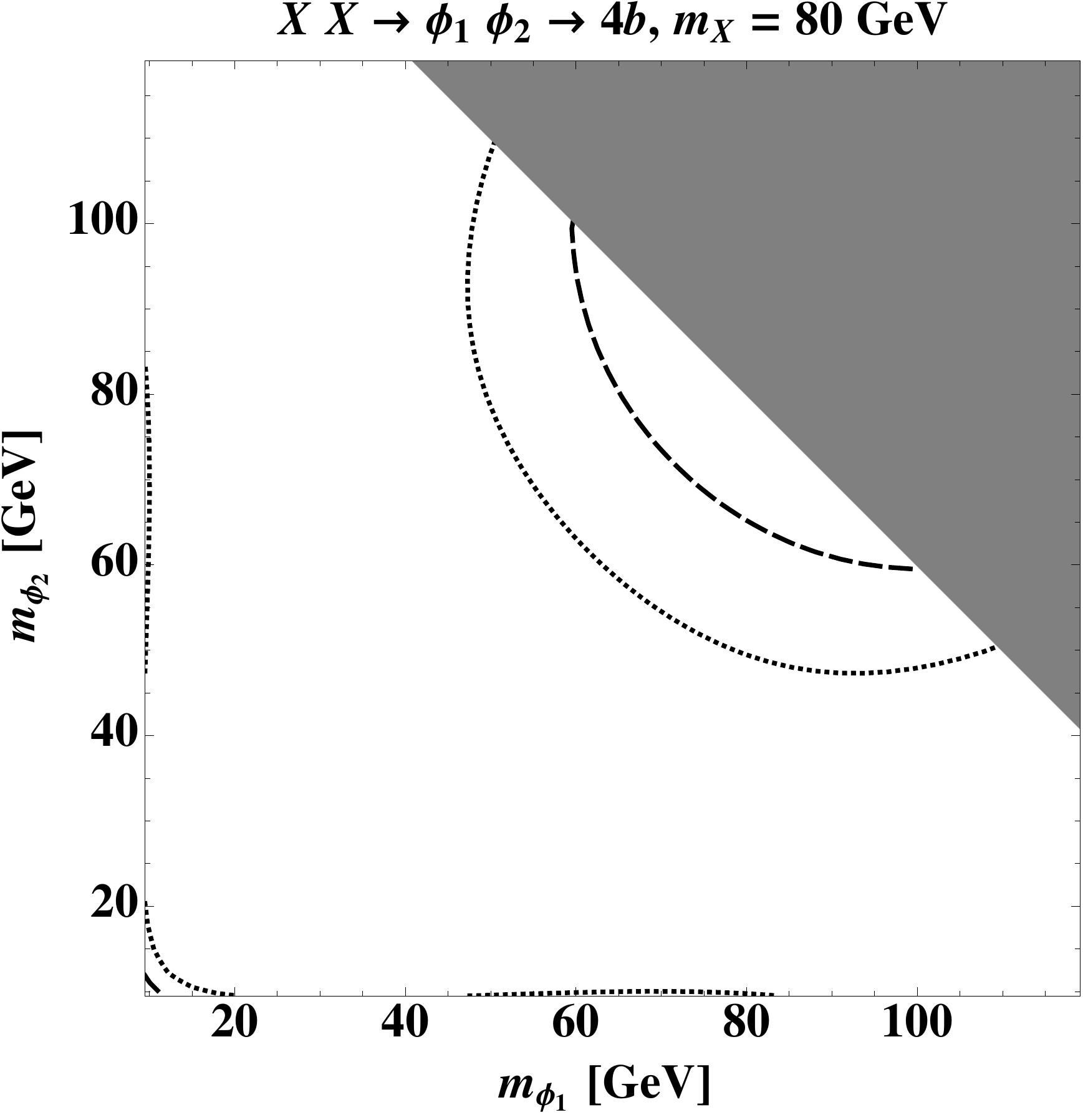} 
\vspace{0.2cm}
\includegraphics[width=0.4\textwidth]{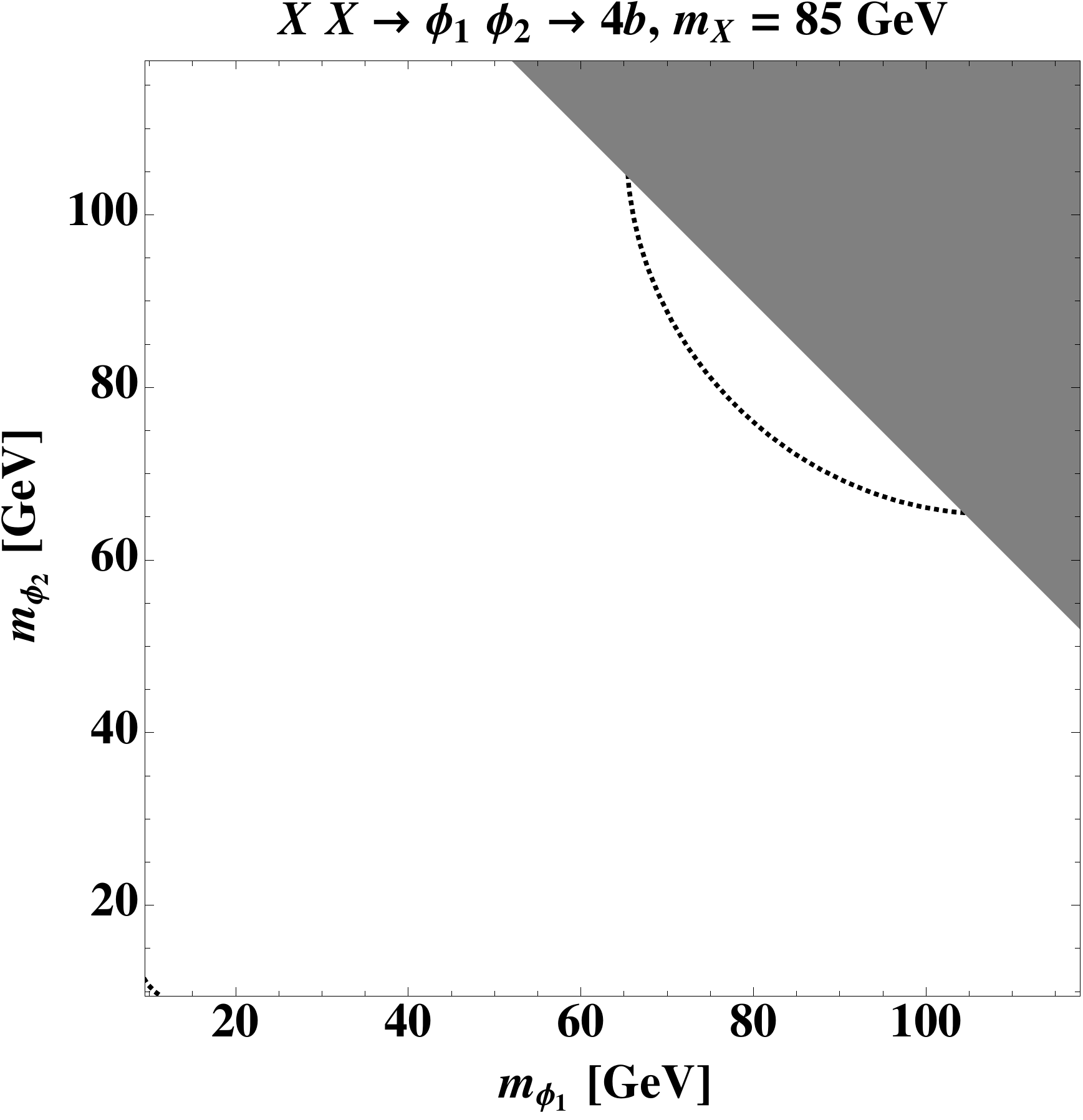}
\caption{The regions of the $m_{\phi_1}-m_{\phi_2}$ plane which lead to a gamma-ray spectrum in agreement with that observed from the Galactic Center, for five different values of the dark matter mass (65, 72, 80 or 85 GeV). The blue dot represents the best-fit point, surrounded by 1, 2 and 3$\sigma$ contours. In these figures, we assume that each $\phi_{1,2}$ decays to $b\bar{b}$. For a wide range of parameters, annihilations through intermediate states can accommodate the gamma-ray spectrum observed from the Galactic Center.}
\label{bbfits}
\end{center}
\end{figure*}

In Fig.~\ref{bbfits}, we show the regions of the $m_{\phi_1}-m_{\phi_2}$ plane that provide a good fit to the spectrum of the gamma-ray excess, for dark matter masses of 65, 72, 80 or 85 GeV. For a wide range of these parameters, the observed spectral shape can be accommodated. The best fits are generally found for dark matter masses in the range of 60-80 GeV and for intermediate particles that are either produced nearly at rest ($m_{\phi_1}+m_{\phi_2} \sim 2 m_X$) or that are not much heavier than $2m_b$. In either of these two limits, the gamma-ray spectrum is identical to that predicted for dark matter annihilating directly to $b\bar{b}$. For other intermediate particle masses, the combined boosts of the $\phi$ and its decay products lead to a somewhat broader spectrum that is less capable of fitting the observed gamma-ray excess. The blue star in the $m_X=72$ GeV frame represents the best-fit point (with $\chi^2=27.6$ over 24 degrees-of-freedom), with solid, dashed and dotted contours representing 1, 2 and 3$\sigma$ regions around that point.\footnote{We point out that our preferred regions differ somewhat from those found in Refs.~\cite{Ko:2014gha,Boehm:2014bia,Abdullah:2014lla,Martin:2014sxa} due to our different statistical weighting of the extracted excess. For instance, Ref.~\cite{Abdullah:2014lla} places a uniform 20\% error on all points, which broadens the peak of the excess and allows more boosted final states to achieve a good fit.}

In addition to potentially altering the spectral shape of the gamma-ray signal from the Galactic Center, dark matter annihilating through intermediate states can also lead to an overall suppression of the gamma-ray emission relative to that predicted by models in which the dark matter annihilates directly to Standard Model fermions. In either case, to obtain a thermal relic density equal to the measured dark matter abundance, we require that the dark matter annihilates with a cross section at freeze-out given by $\sigma v \simeq 2.2 \times 10^{-26}$ cm$^3$/s. The power produced through dark matter annihilations, however, is proportional to $\sigma v/m_X$.\footnote{The annihilation rate and power per annihilation scale as $\sigma v/m^2_X$ and $m_X$, respectively.}  As a result, the higher dark matter masses required in the case of cascade annihilations reduces the intensity of the predicted gamma-ray signal.

We also point out that if the intermediate particles are nearly degenerate in mass to the dark matter, this can lead to a phase space suppression of the annihilation cross section that is more pronounced in the Galaxy today than it was at the time and temperature of thermal freeze-out, reducing the annihilation rate in the Galactic Center by a factor of:
\begin{equation} \label{rel-enh}
\frac{\langle \sigma v \rangle_{\rm today}}{\langle \sigma v \rangle_{\rm freeze-out}} \simeq \left[ \frac{\epsilon+v_{0}^2(1-\epsilon)}{\epsilon+v_{\rm FO}^2(1-\epsilon)} \right]^{k/2},
\end{equation}
where $v_{\rm FO} \simeq 0.3$, $v_0 \simeq 10^{-3}$, $\epsilon \equiv (m_X^2-m_\phi^2)/m_X^2$, and $k=1(3)$ for annihilation to two scalars (vectors). For a mass splitting of order 1\% (5\%), the present-day annihilation rate will be suppressed by a factor of a few (a few percent). 

While these factors impacting the normalization of the gamma-ray signal are not insignificant, they can be compensated by adjusting the mass of the Milky Way's dark matter profile, which is uncertain at the level of a factor of a few~\cite{Iocco:2011jz}.

\section{A Hidden Photon Model}
\label{hiddenphoton}

In this section, we consider a simple model in which the dark matter, $X$, is a Dirac fermion charged under a new $U(1)_X$. This gauge group is broken by some dark Higgs field, which provides a massive vector boson, $\phi$, sometimes called a hidden or dark photon. Together, the dark matter and vector boson reside within a hidden sector, with no direct couplings to the Standard Model. Dark matter interacting through hidden sector forces has been widely discussed within a variety of contexts~\cite{ArkaniHamed:2008qn,Pospelov:2008jd,Cholis:2008qq,Kang:2010mh,Finkbeiner:2007kk,Cholis:2008vb,Pospelov:2007mp,Morrissey:2009ur,Chang:2010yk,Cohen:2010kn,Meade:2009rb,Hooper:2012cw,Essig:2013goa}.

\begin{figure}[b]
\begin{center}
\includegraphics[width=0.4\textwidth]{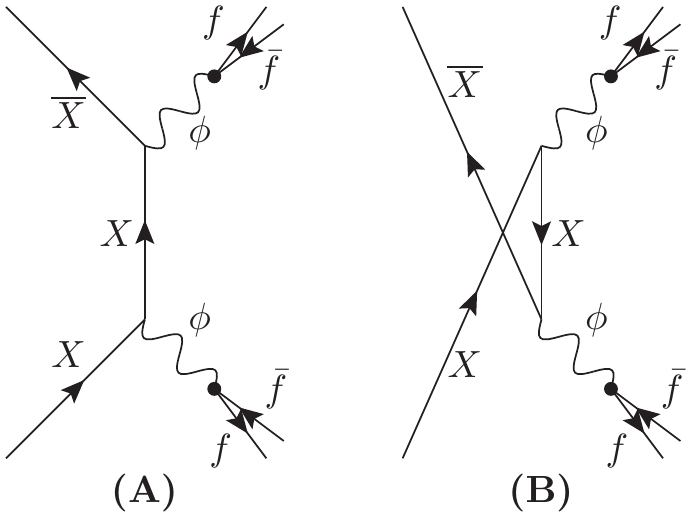} 
\caption{Annihilation of dark matter into two hidden photons via {\bf (A)} $t$- and {\bf (B)} $u$-channel diagrams. The hidden photons decay into Standard Model particles through kinetic mixing with the Standard Model photon.}
\label{fd-U1d}
\end{center}
\end{figure}

\begin{figure*}[t]
\begin{center}
\includegraphics[width=0.6\textwidth]{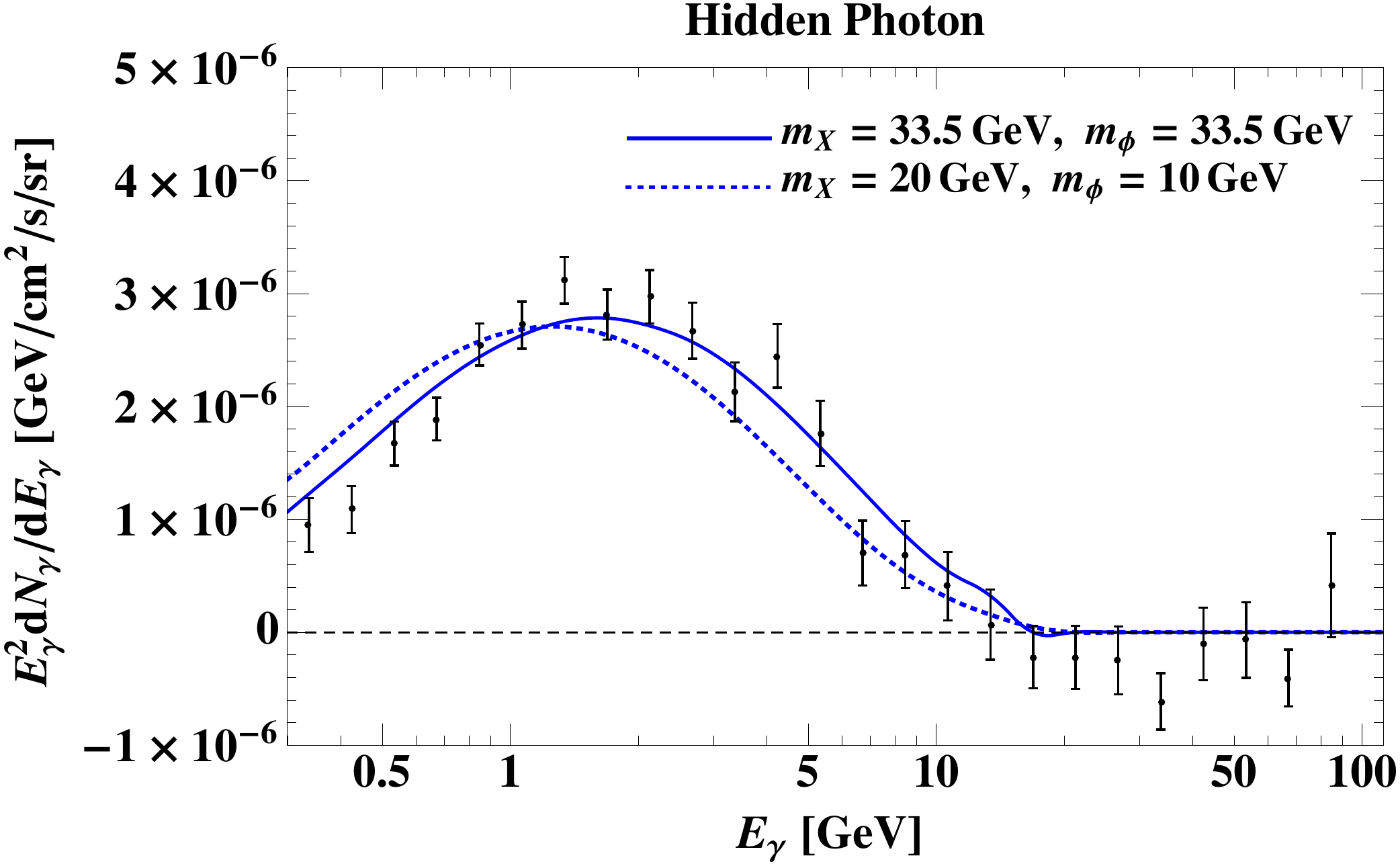}
\caption{The shape of the gamma-ray spectrum produced by the annihilations of dark matter in the hidden photon model described in Sec.~\ref{hiddenphoton}, for two choices of parameters. This is compared to the spectrum of the observed gamma-ray excess, as reported in Ref.~\cite{Daylan:2014rsa}.}
\label{hiddenphotonspec}
\end{center}
\end{figure*}

If the hidden photon is lighter than the dark matter candidate, then dark matter annihilations will be dominated by the $t$- and $u$-channel exchange of an $X$ into a pair of $\phi$ particles, as shown in Fig.~\ref{fd-U1d}. The cross section for this process is fully determined by the masses $m_X$ and $m_\phi$, and the $U(1)_X$ charge, $g_X$, and is given by: 
\begin{eqnarray}
\label{U1d-ann}
\langle \sigma v \rangle_{XX \rightarrow \phi \phi} &\simeq& \frac{ \pi \alpha^2_X}{m^2_{X}} \, \frac{(1-m^2_\phi/m^2_X)^{3/2}}{(1-m_\phi^2/2m_X^2)^2} \\
&\simeq& \, \, 2.2 \times 10^{-26} \, {\rm cm}^3/{\rm s}  \nonumber \\ ~~~&\times& \bigg(\frac{g_X}{0.1}\bigg)^4 \, \bigg(\frac{34\,{\rm GeV}}{m_X}\bigg)^2  \, \frac{(1-m^2_\phi/m^2_X)^{3/2}}{( 1-m_\phi^2/2m_X^2 )^2}, \nonumber
\end{eqnarray}
where $\alpha_X \equiv g^2_X/4\pi$ is the fine structure constant of $U(1)_X$. Throughout the remainder of this section, we will set $g_X$ such that $\sigma v = 2.2 \times 10^{-26}$ cm$^3$/s, thus generating a thermal relic abundance in agreement with the cosmological dark matter density~\cite{Steigman:2012nb}. This cross section also leads to a gamma-ray signal that, within uncertainties in the normalization of the Milky Way's dark matter halo profile, is in agreement with that observed from the Galactic Center~\cite{Daylan:2014rsa}. 

The size of the coupling, $g_X$, has no direct implication for the strength with which the dark matter couples to the Standard Model. If the photon and the $\phi$ undergo kinetic mixing, however, this can induce a coupling between the hidden sector and the Standard Model (alternatively, one could also consider mixing between the $\phi$ and the $Z$). This kinetic mixing can be described by a Lagrangian of the form $\mathcal{L}=\frac{1}{2}\epsilon F'_{\mu \nu} F^{\mu \nu}$~\cite{Holdom:1985ag}, which is allowed by all symmetries of the theory. Kinetic mixing with the photon then allows for suppressed couplings between the $\phi$ and the particles of the Standard Model, proportional to their electric charge. Although there is no robust prediction for the size of this coupling (any value is technically natural~\cite{ArkaniHamed:2008qp}), arguments can be made in support of some values. For example, if the Standard Model is embedded within a Grand Unified Theory (GUT), a non-zero value of $\epsilon$ can only be generated after GUT breaking at the loop level. Such a loop of heavy states carrying both hypercharge and $X$ gauge charge naturally leads to kinetic mixing of the following order~\cite{Holdom:1985ag,Baumgart:2009tn,Cohen:2010kn}:
\begin{align}
\begin{split}
\label{epsilon}
\epsilon &\sim \frac{g_X g_Y \cos \theta_W}{16 \pi^2} \, \ln\left(\frac{M'^2}{M^2}\right) \\
&\sim 2 \times 10^{-4} \, \bigg(\frac{g_X}{0.1}\bigg) \, \ln\bigg(\frac{M'^2}{M^2}\bigg), 
\end{split}
\end{align}
where $M'$ and $M$ are the masses of the particles in the loop. Thus we expect the kinetic mixing to occur at a level of $\epsilon\sim 10^{-3}$ or less, modulo the possibility of a large hierarchy between $M'$ and $M$. If the splitting between the different components of the GUT multiplet is instead generated at loop order, then $\epsilon$ will be suppressed by two loops, further reducing the expected value of $\epsilon$. Throughout this section, we will assume that $\epsilon$ is large enough to have kept the hidden sector in thermal equilibrium with the Standard Model throughout the process of dark matter freeze-out. In particular, for values of $\epsilon \gsim 10^{-7}$, the rate of $f \gamma \leftrightarrow f \phi$ is sufficient to ensure that the system will be thermalized before the temperature of decoupling.

The gamma-ray spectrum from dark matter annihilations in this model depends on the dominant decay channels of the $\phi$. For $m_{\phi}$ greater than a few GeV, the $\phi$ decays directly to pairs of quarks and charged leptons. Since these decays are mediated by the Standard Model photon, the branching fractions are determined only by their electric charge and phase space factors. In Fig.~\ref{hiddenphotonspec} we show examples of the gamma-ray spectrum from dark matter annihilation in this model. As noted above, we see that producing the $\phi$'s near rest ($m_{\phi} \sim m_X$) yields the best-fit. Much lighter hidden photons lead to a broader spectrum, in some conflict with the shape of the observed gamma-ray excess. Small mass splittings within the hidden sector are not difficult to achieve, and can be realized in a variety of concrete models~\cite{Morrissey:2009ur,Essig:2013goa,Ruderman:2009tj,Kearney:2013xwa,Fan:2012gr}.

\begin{figure}[t]
\begin{center}
\includegraphics[width=0.4\textwidth]{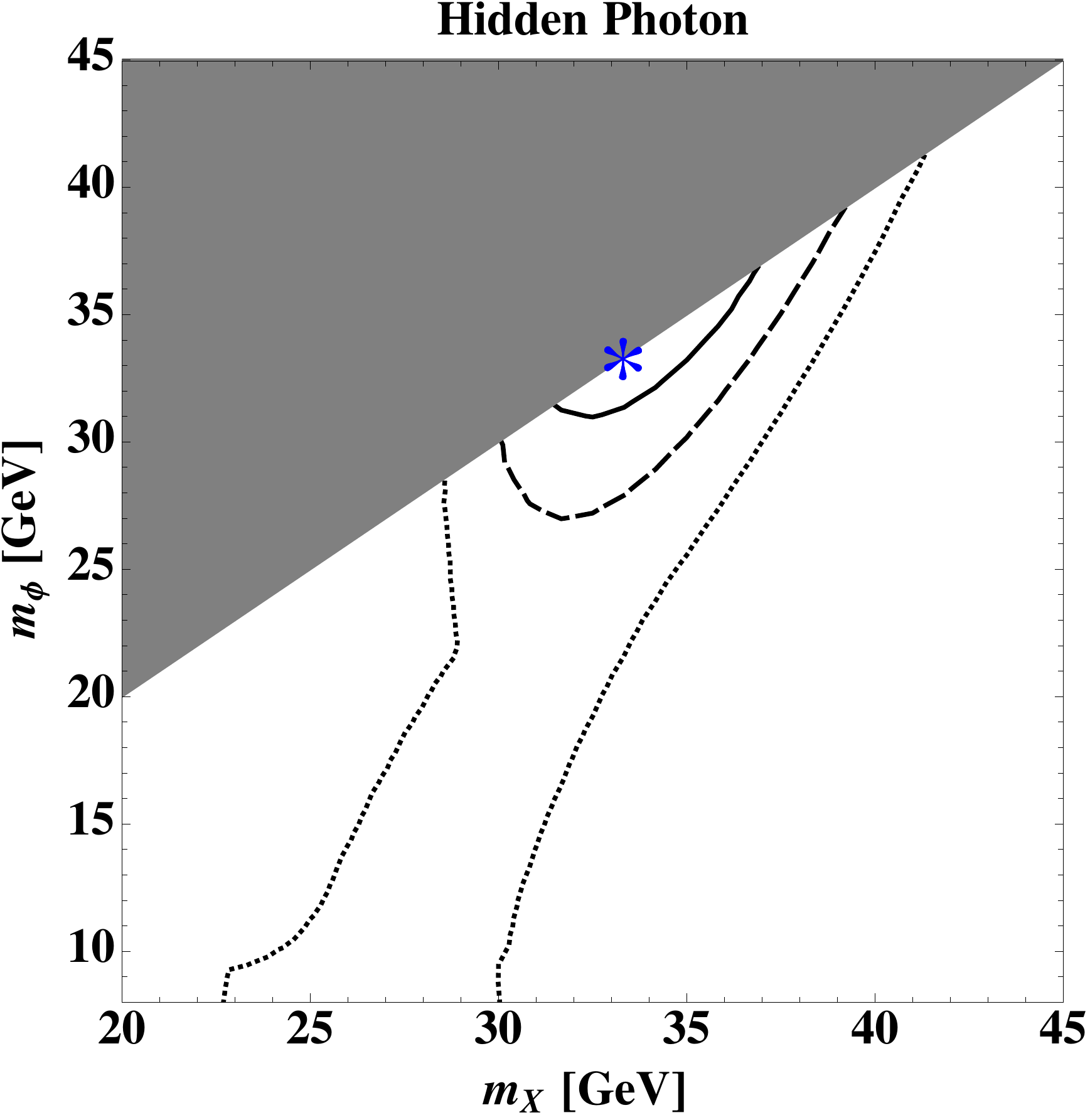}
\caption{The regions of the parameter space in the hidden photon model that provide a good fit to the spectral shape of the gamma-ray excess. The blue dot represents the best-fit point, and is surrounded by 1, 2 and 3$\sigma$ contours.}
\label{U1d-plot}
\end{center}
\end{figure}

In Fig.~\ref{U1d-plot}, we show the regions of the $m_X-m_\phi$ plane that are capable of providing a good fit to the observed Galactic Center gamma-ray excess. The best-fit point (shown as a blue star) provides a reasonable fit to the data, corresponding to $\chi^2=34.9$ over 24 degrees-of-freedom. At the 2$\sigma$ level, there is a strong preference for $m_X \simeq m_\phi$, with $30\,{\rm GeV} \lsim m_X \lsim 40\,{\rm GeV}$. At 3$\sigma$, lower values of $m_{\phi}$ are also allowed.  After setting the annihilation cross section to the value required to generate the desired relic abundance ($\sigma v \simeq 2.2 \times 10^{-26}$ cm$^3$/s), we find that the overall normalization of the gamma-ray excess can be accommodated for local dark matter density of $\rho_{\rm local} \simeq 0.3$ GeV/cm$^3$, in good agreement with dynamical measurements~\cite{Iocco:2011jz}.

Although interactions between the hidden sector and the Standard Model are suppressed in this model, kinetic mixing between the $\phi$ and the photon leads to vector-mediated spin-independent elastic scattering between the dark matter and protons. The cross section for this process is given by:
\begin{eqnarray} \label{DDU1d}
\sigma_{Xp} &=&16\pi \, \alpha_{\rm EM} \, \alpha_X \, \epsilon^2 \, \frac{ \mu_{Xp}^2}{m^4_{\phi}} \\
&\simeq& 1.2 \times 10^{-45} \, {\rm cm}^2 \, \bigg(\frac{\epsilon}{10^{-4}}\bigg)^2 \bigg(\frac{g_X}{0.1}\bigg)^2 \bigg(\frac{30\,{\rm GeV}}{m_{\phi}}\bigg)^4, \nonumber
\end{eqnarray}
where $\alpha_{\rm EM}$ is the electromagnetic fine structure constant and $\mu_{Xp}$ is the dark matter-proton reduced mass. For dark matter which scatters equally with protons and neutrons, the LUX experiment requires $\sigma_{Xp} \lsim 8 \times 10^{-46}$ cm$^2$, for $30\,{\rm GeV} \lsim m_X \lsim50$ GeV. As elastic scattering is mediated by the photon in this model, however, the dark matter does not scatter with neutrons. Hence, the effective nucleon level cross section with a xenon nucleus is reduced by a factor of $A^2/Z^2 \sim 5.8$, corresponding to a rescaled bound $\sigma_{Xp} \lsim 4.6 \times 10^{-45}$ cm$^2$. Comparing this to the result of Eq.~\ref{DDU1d}, we find that LUX is currently sensitive to values of $\epsilon$ greater than approximately $10^{-4}$, near the range of values suggested by Eq.~\ref{epsilon}.  Consequently, if neither LUX nor XENON1T observes a signal within the next few years, that would disfavor models in which kinetic mixing between the photon and hidden photon is generated at the one-loop level.

As the hidden photon decays to $e^+ e^-\sim$15\% of the time in this model, dark matter annihilations taking place in the local halo of the Milky Way are predicted to induce a spectral feature in the cosmic ray positron fraction. The lack of such a feature in the spectrum reported by AMS can be used to place a constraint on this model \cite{Kong:2014haa}. For $m_{\phi} \simeq m_{X}$ and for reasonable estimates of the local density and propagation parameters, we arrive at $\sigma v \lsim 1.1\times 10^{-26}$ cm$^3$/s~\cite{Bergstrom:2013jra}. And although this constraint is in tension with the value required from relic density considerations, a local under-density of dark matter or a higher local energy loss rate for electrons could plausibly reconcile this model with AMS.

\section{A Hidden Sector within the Generalized NMSSM}
\label{nmssm}

In this section, we consider a supersymmetric model that includes a sector that is largely sequestered from the Standard Model and its superpartners. This is naturally realized within the context of the generalized Next-to-Minimal Supersymmetric Standard Model (NMSSM), in which the Higgs singlet and its superpartner, the singlino, couple to Standard Model fields only through small mixing angles.  We use the term ``generalized" to indicate that we impose no additional $\mathbb{Z}_N$ symmetries, as are sometimes implicitly included in such models. For other recent work on dark matter in the NMSSM, with various choices of additional symmetries and target phenomenology, see Refs.~\cite{Ellwanger:2009dp,Ross:2012nr,Das:2010ww,Draper:2010ew,Carena:2011jy,Kozaczuk:2013spa,Gunion:2005rw,Belanger:2005kh,Cerdeno:2007sn,Hooper:2009gm}.

We begin by writing down the general superpotential and soft Lagrangian for the generalized NMSSM:
\begin{eqnarray}
W^\text{Higgs} &=& (\mu + \lambda \hat{S}) \hat{H}_u \hat{H}_d + \xi_F \hat{S} + \frac{1}{2} \mu^\prime \hat{S}^2 + \frac{1}{3} \kappa \hat{S}^3~~~
\\
-\mathcal{L}_\text{soft}^\text{Higgs} &=& m_{H_u}^2 | H_u |^2 + m_{H_d}^2 |H_d|^2 + m_S^2 |S|^2 \\  \nonumber
&+& \Big[ (B \mu + A_\lambda \lambda S) H_u H_d + \xi_S S  \\ \nonumber
&+&\frac{1}{2} B^\prime \mu^\prime S^2 + \frac{1}{3} A_\kappa \kappa S^3 + h.c. \Big], 
\end{eqnarray}
where $S$ represents the additional singlet scalar, and hats denote superfields. Although we are allowed to make a field redefinition to shift away {\it one} of the dimensionful parameters (typically the tadpole coefficent, $\xi_F$), we retain all such terms here for the sake of generality.

The neutralino mass matrix in the $\tilde{B}-\tilde{W}^0-\tilde{H}_d-\tilde{H}_u-\tilde{S}$ basis is given by~\cite{Ellwanger:2009dp}:
\begin{widetext}
\begin{equation}
{\cal M}_{\tilde \chi^0} = \begin{pmatrix} M_1 & 0 & -g_1 v_d/\sqrt{2} & g_1 v_u/\sqrt{2} & 0 \\ 
0 & M_2 & g_2 v_d/\sqrt{2} & -g_2 v_u/\sqrt{2} & 0 \\ 
-g_1 v_d/\sqrt{2} & g_2 v_d/\sqrt{2} & 0 & -(\mu+\lambda v_s) & -\lambda v_u \\
g_1 v_u/\sqrt{2} & -g_2 v_u/\sqrt{2} & -(\mu+\lambda v_s) & 0 &  -\lambda v_d \\
0 & 0 & -\lambda v_u & -\lambda v_d & 2 \kappa v_s+\mu' \end{pmatrix},
\end{equation}
\end{widetext}
where $M_1$ and $M_2$ are the bino and neutral wino masses, $v_u$ and $v_d$ are the vacuum expectation values of the Higgs doublet for up- and down-type fermions, respectively, and $v_s$ is the vacuum expectation value of the Higgs singlet. We will focus on the case in which the lightest supersymmetric particle (stabilized by $R$-parity) is a highly singlino-like neutralino, with a mass given by $m_\chi \simeq 2 \kappa v_s+\mu'$.  This can be realized when $\lambda \ll 1$, and $M_1,M_2,\mu \gg 2 \kappa v_s+\mu'$.

\begin{figure}[t]
\begin{center}
\includegraphics[width=0.4\textwidth]{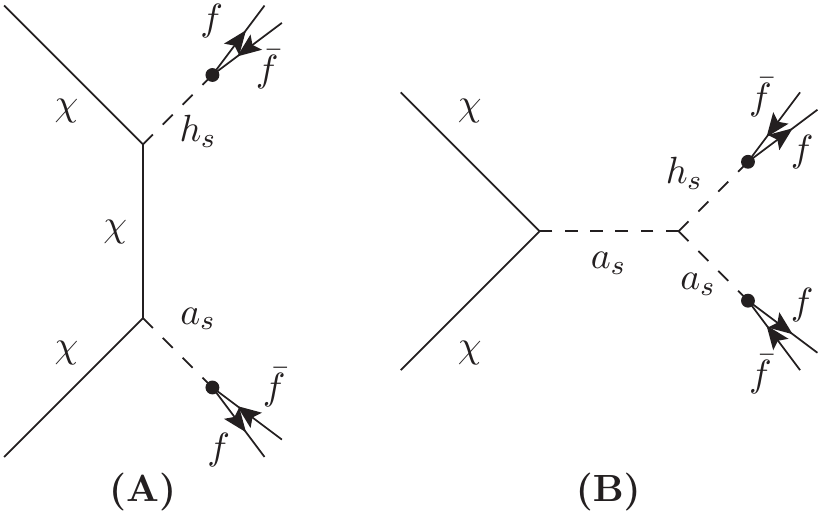} 
\caption{Annihilation of singlino-like neutralino dark matter into a higgs singlet scalar ($h_s$) and pseudoscalar ($a_s$) via {\bf (A)} $t$- and {\bf (B)} $s$-channel diagrams. The $h_s$ and $a_s$ each decay into Standard Model fermions via mass mixing with the Higgs bosons of the MSSM. The $u$-channel diagram is not shown.}
\label{fd-NMSSM}
\end{center}
\end{figure}

$S$ is a complex scalar and a gauge singlet. After settling into the electroweak vacuum, it gets a vacuum expectation value and manifests as two physical states; we take the convention $S=v_s+(h_s+i a_s)/\sqrt2$. The states $h_s$ and $a_s$ mix with the scalar and pseudoscalar neutral Higgs bosons of the MSSM, respectively. We assume that there is no significant CP violation in the NMSSM scalar sector so that the mixing factorizes. In the $(h_{u},h_{d},h_s)$ basis, the components of the CP-even mass squared matrix are given by:
\begin{eqnarray}\label{scalarmass}
M_{S,11}^2 &=&  \Big[B \mu + \lambda v_s ( A_\lambda + \mu^\prime+\kappa v_s)+\lambda \xi_F \Big]\cot{\beta} \nonumber \\ &+& M_Z^2 \sin^2{\beta}
\nonumber \\
M_{S,12}^2 &=&  - \big[B \mu + \lambda v_s (A_\lambda+\mu^\prime+\kappa  v_s) + \lambda \xi_F \big]  \nonumber \\\nonumber
&+& \frac{1}{2} \left(2 \lambda ^2 v^2 - M_Z^2 \right) \sin{2 \beta}
\nonumber \\
M_{S,13}^2 &=& \lambda  v \Big[ 2 (\mu +\lambda v_s)  \sin{\beta} - (A_\lambda+\mu^\prime+2 \kappa  v_s)\cos{\beta} \Big]
\nonumber \\
M_{S,22}^2 &=&  \Big[B \mu + \lambda v_s (A_\lambda+\mu^\prime+\kappa v_s)+\lambda \xi_F \Big] \tan{\beta}  \nonumber \\ &+& M_Z^2 \cos^2{\beta}
\nonumber \\
M_{S,23}^2 &=& \lambda  v \Big[ 2 (\mu +\lambda v_s) \cos{\beta} - (A_\lambda+\mu^\prime+2 \kappa v_s) \sin{\beta} \Big]
\nonumber \\
M_{S,33}^2 &=&  \kappa  v_s (A_\kappa+3 \mu^\prime + 4 \kappa  v_s )  + \frac{1}{2} \lambda  \frac{v^2}{v_s} (A_\lambda+\mu^\prime) \sin{2 \beta}  \nonumber \\
&-&  \frac{1}{v_s} ( \mu^\prime \xi_F +  \xi_S + \lambda  \mu  v^2 ). 
\end{eqnarray}
After rotating the basis and dropping the Goldstone mode, the CP-odd mass squared matrix in the $(A,a_s)$ basis is given by:
\begin{eqnarray} \label{pseudomass}
M_{P,11}^2 &=& 2 \Big[  B \mu + \lambda v_s (A_\lambda + \mu^\prime + \kappa v_s) + \lambda \xi_F \Big] \frac{1}{\sin{2\beta}}
\nonumber \\
M_{P,12}^2 &=& \lambda  v (A_\lambda-\mu^\prime-2 \kappa v_s) 
\nonumber \\
M_{P,22}^2 &=& \frac{1}{2} \lambda  \frac{v^2}{v_s}  (A_\lambda + \mu^\prime + 4 \kappa  v_s )\sin{2 \beta}- \kappa v_s (3A_\kappa + \mu^\prime ) \nonumber \\
&-& 2 B^\prime \mu^\prime - 4 \kappa \xi_F -\frac{1}{v_s} ( \mu^\prime \xi_F+ \xi_S + \lambda  \mu  v^2 ).
\end{eqnarray}
We take the alignment limit ($\beta = \alpha+\pi/2$) so that the Higgs boson discovered at the LHC is expected to be very Standard Model like.

As was the case for the neutralinos, the scalar singlet-sector particles decouple from the MSSM for small values of $\lambda$. In the limit of small $\lambda$, the CP-even and CP-odd mass eigenstates $h_s, a_s$ have masses approximately given by the square roots of the $33$ and $22$ entries in Eqs.~\ref{scalarmass} and~\ref{pseudomass}, respectively. We point out that all of the terms not proportional to $\lambda$ in the 22 entry of Eq.~\ref{pseudomass} are negative. Since we are assuming that $\lambda$ is very small in order to suppress the off-diagonal entries, we have to assume that $B'$ is large and negative to prevent a tachyonic $a_s$. Since $B'$ does not enter the other mass matrices, we have the parameter freedom to tune $B'$ as needed. Since $A_\kappa$ controls $m_{h_s}^2$ but does not enter ${\cal M}_{\tilde \chi^0}$, this further implies that $m_{a_s},m_{h_s}$, and $m_\chi$ are effectively independent and observe no special mass relations.

Assuming that the sum of the singlet-like scalar and pseudoscalar Higgs boson masses is smaller than twice the singlino mass ($m_{h_s}+m_{a_s} < 2 m_{\chi}$), dark matter annihilations will proceed dominantly to the $a_s h_s$ final state \cite{Nomura:2008ru,Mardon:2009rc} through a combination of $t/u$-channel singlino exchange and $s$-channel $a_s$ exchange diagrams, as shown in Fig.~\ref{fd-NMSSM}. In the low-velocity limit, the cross section for this process is given by~\cite{Hooper:2009gm}:
\begin{eqnarray}
\label{ha}
&&\langle \sigma v \rangle_{\chi \chi \rightarrow a_s h_s} \simeq \frac{\kappa^4}{4 \pi m^2_{\chi}}  \, v_{\rm out}\\
&\times& \bigg( \frac{4m_{\chi}^2 + m_{a_s}^2 - m_{h_s}^2}{4m_{\chi}^2 - m_{a_s}^2 - m_{h_s}^2} - \frac{(2 \kappa v_s + \mu' - A_{\kappa} ) \, m_\chi}{4m_{\chi}^2-m_{a_s}^2}  \bigg)^2 \nonumber,
\end{eqnarray}
where 
\begin{equation}
v_{\rm out} =\left[\left(1-\frac{(m_{a_s}+m_{h_s})^2}{4m_\chi^2} \right) \left(1-\frac{(m_{a_s}-m_{h_s})^2}{4m_\chi^2} \right) \right]^{1/2}  .
\end{equation}
Although singlinos can also annihilate into $h_s h_s$ and/or $a_s a_s$ final states, these processes are additionally suppressed by two powers of velocity. In the case that annihilations proceed largely through the first term in Eq.~\ref{ha}, corresponding to the $t/u$-channel process, the cross section yields:
\begin{eqnarray}
\langle \sigma v \rangle_{\chi \chi \rightarrow a_s h_s} &\sim& 2.2 \times 10^{-26} \, {\rm cm}^3/{\rm s} \,  \,   \nonumber 
\\  &\times& \bigg(\frac{\kappa}{0.10}\bigg)^4 \, \bigg(\frac{m_\chi}{67 \, {\rm GeV}}\bigg)^{-2} v_{\rm out}.
\end{eqnarray}

\begin{figure*}[t]
\begin{center}
\includegraphics[width=0.6\textwidth]{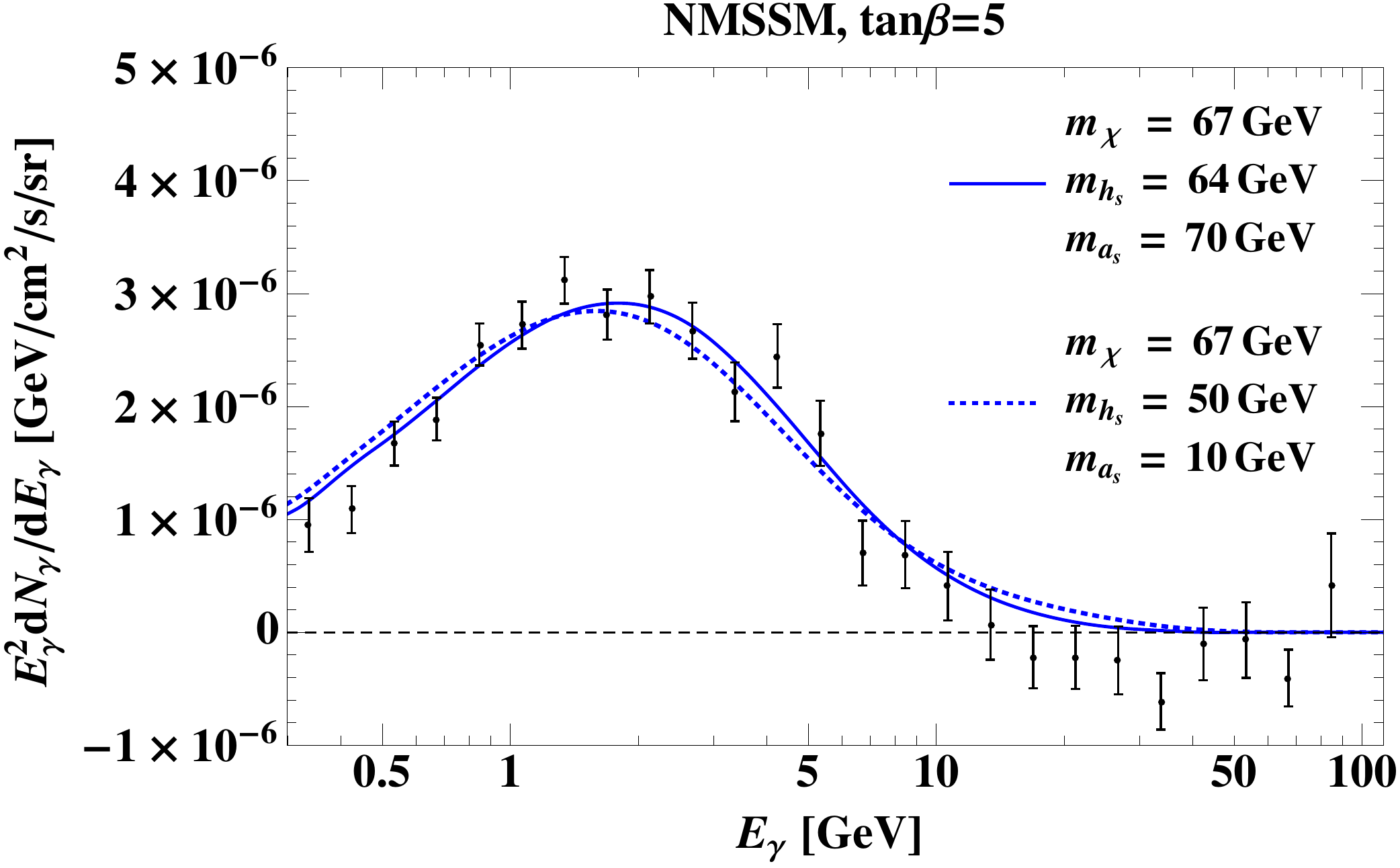}
\caption{The shape of the gamma-ray spectrum produced by the annihilations of singlino dark matter in the generalized NMSSM, as described in Sec.~\ref{nmssm}, for two choices of parameters. This is compared to the spectrum of the observed gamma-ray excess, as reported in Ref.~\cite{Daylan:2014rsa}.}
\label{nmssmspec}
\end{center}
\end{figure*}

\begin{figure*}[!t]
\begin{center}
\includegraphics[width=0.37\textwidth]{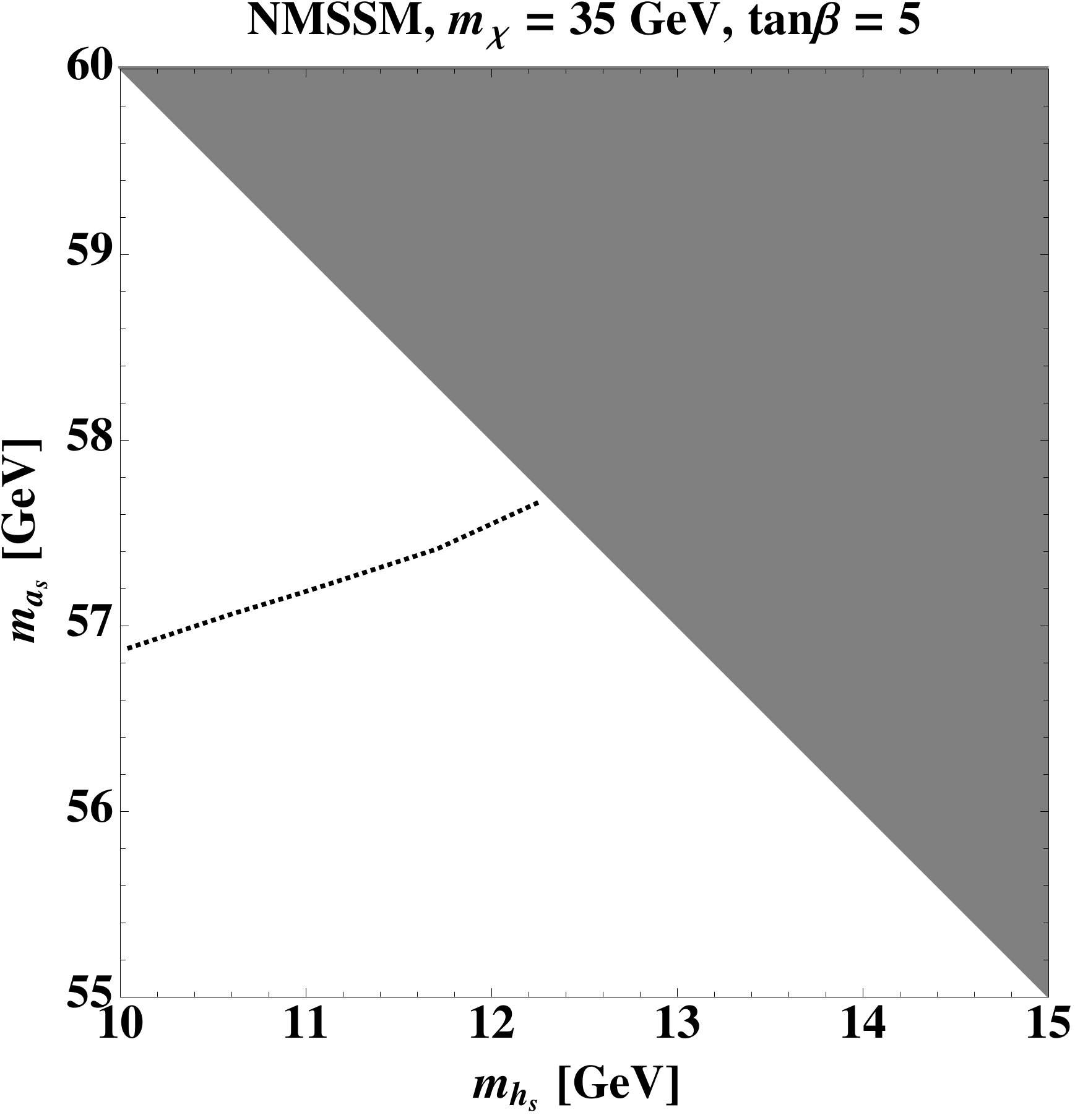}
\hspace{0.15cm}
\includegraphics[width=0.37\textwidth]{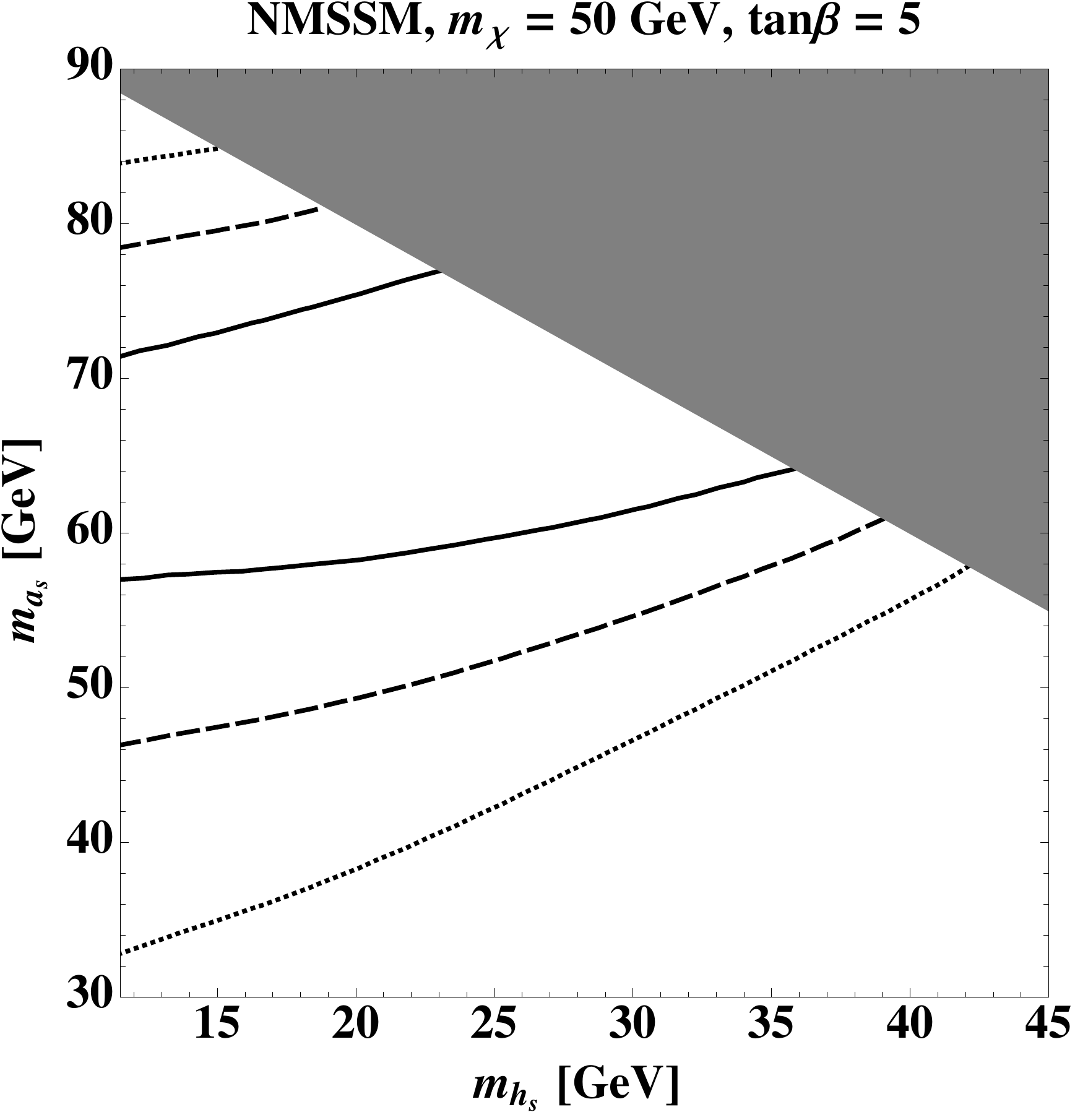} \\
\vspace{0.15cm}
\includegraphics[width=0.37\textwidth]{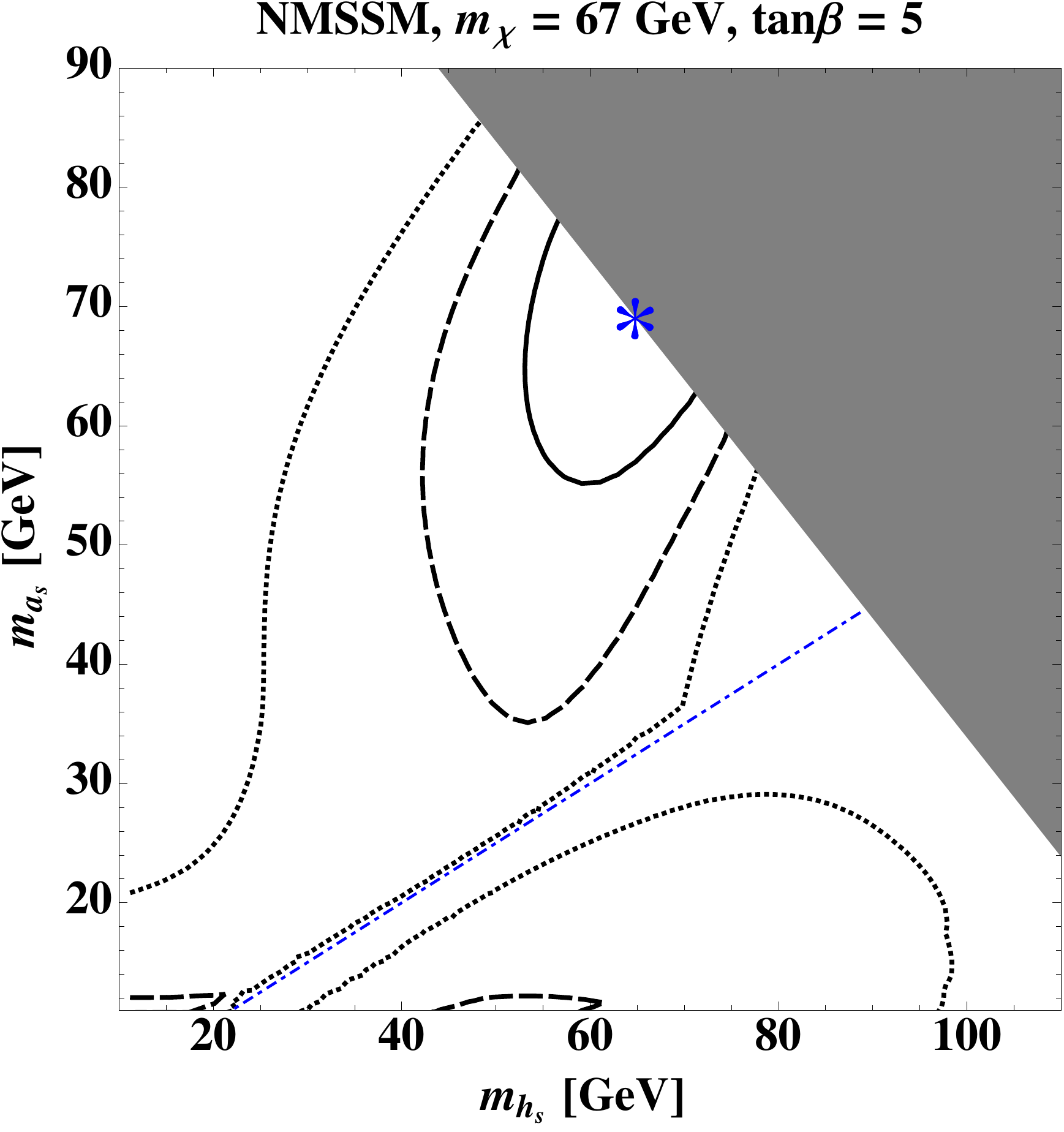}
\hspace{0.15cm}
\includegraphics[width=0.37\textwidth]{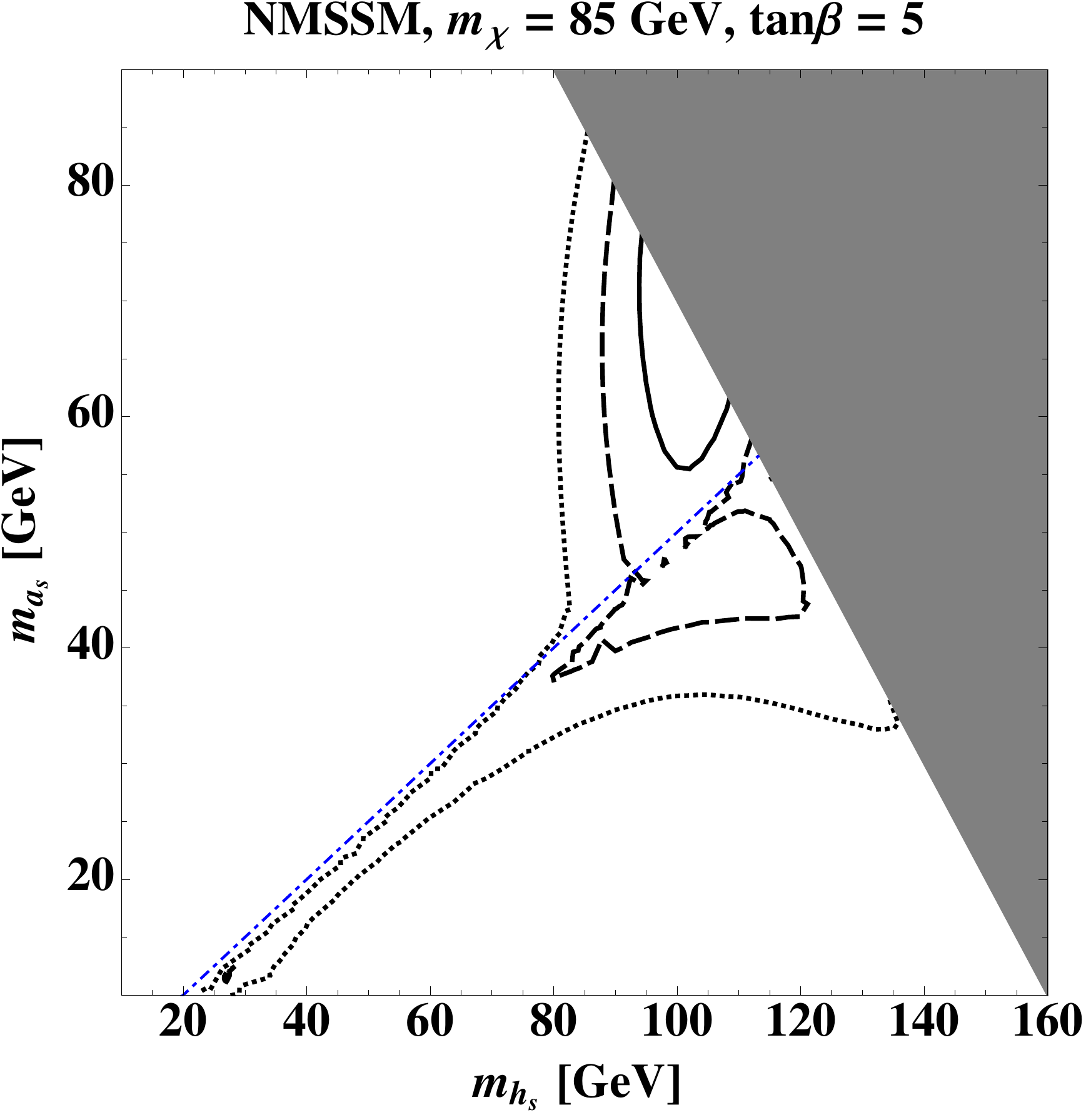}
\caption{The regions of the parameter space in the generalized NMSSM that provide a good fit to the spectral shape of the gamma-ray excess. The blue dot represents the best-fit point, and is surrounded by 1, 2 and 3$\sigma$ contours. Below the blue dot-dashed lines in the bottom two figures, the decay $h_s \to a_s a_s$ is kinematically accessible.}
\label{nmssmfit}
\end{center}
\end{figure*}

After annihilation to $a_s h_s$, these particles decay to Standard Model fermions with branching ratios proportional to mass, and thus are typically dominated by the heaviest kinematically available quarks or leptons.  Other decays are possible in extreme ranges of parameter space, however. For instance, the branching ratio for $h_s \to a_s a_s$ is expected to be large if $m_{h_s} > 2 m_{a_s}$. Alternatively, if $m_{h_s} > m_{a_s}+m_Z$, one might expect the $h_s$ to decay into a $a_sZ$ final state. This coupling, however, is suppressed by $\cos (\beta-\alpha)$ and is negligible in the limit under consideration~\cite{Djouadi:2005gj}.

In Fig.~\ref{nmssmspec}, we plot the gamma-ray spectrum from singlino annihilation, for two choices of parameters. In Fig.~\ref{nmssmfit}, we show the regions of the parameter space which allow for a good fit to the gamma-ray excess, for four choices of the singlino mass: $m_\chi=$ 35, 50, 67, and 85 GeV\footnote{In both of these figures, we take $\tan \beta = 5$. Taking $1 \lsim \tan \beta \lsim 10$ will not qualitatively alter our conclusions.}. In this case, the best-fit (shown as a blue star) provides a good-fit to the data, corresponding to $\chi^2=30.0$ over 24 degrees-of-freedom. As noted above, the coupling of both the $a_s$ and the $h_s$ to Standard Model fermions is proportional to the fermion mass and so both go dominantly to $b \bar b$ pairs. Moreover, the $a_s$ inherits an additional $\tan \beta$ enhancement of its couplings to down-type quarks, and thus decays almost exclusively to $b \bar b$ pairs. In contrast to the roughly democratic spectrum favored by the hidden photon scenario, where many light fermions are produced, the NMSSM final state spectrum strongly favors the heaviest accessible particle, with the additional $\tan \beta$ enhancement.

As is visible in the $m_\chi=67$ and 85 GeV frames of Fig.~\ref{nmssmfit}, the 2$\sigma$ and 3$\sigma$ contours are truncated near the boundary where $h_s \to a_s a_s$ becomes kinematically accessible. Near this threshold, the $a_s$'s from the $h_s$ decay produce relatively soft $b \bar b$ pairs. For large values of $m_{h_s}$, this leads to a relatively broad gamma-ray spectrum, and does not provide a good-fit to the gamma-ray excess. For $m_{h_s} \lsim m_{\chi}$, however, acceptable fits can be obtained.

As long as $m_{h_s} \ll m_h,m_H$, elastic scattering with nuclei is dominated by $h_s$ exchange. The coupling to quarks is given by a small mixing angle with the light MSSM-like Higgs, $h$. For this process, the cross section for scattering off nucleons is given by:
\begin{eqnarray}
\sigma_{\chi N} &\simeq& \frac{\kappa^2 \mu^2_{\chi n} m_n^2}{4 \pi v^2 m^4_{h_s}} \bigg[\sum_{q=u,d,s} f_{T_q}+ \frac29 f_{TG} \bigg]^2 \sin^2 \theta
\\&\simeq& 3.2 \times 10^{-46}\,{\rm cm}^2 \left(\frac\kappa{0.10} \right)^2  \left(\frac{m_{h_s}}{67\,{\rm GeV}} \right)^{-4} \nonumber
\\&&~~~~~~~~~~\times  \left(\frac\lambda{10^{-3}} \right)^2   \left(\frac\mu{2\,{\rm TeV}} \right)^2 ,
\end{eqnarray}
where $\theta$ is the mixing angle between $h$ and $h_s$. In the limit of $\mu \gg \lambda v_s$ and $(A_{\lambda}+\mu'+2\kappa v_s)/\tan\beta$, the diagonalization of Eq.~\ref{scalarmass} yields $\sin \theta \simeq 2\lambda v \mu/m^2_h$. In order for this cross section to evade exceeding the current constraint from LUX, we must require $\lambda \lsim 10^{-3} \times (2 \, {\rm TeV}/\mu)$. Although small, a coupling of this size is not unnatural. For instance, if there is an erstwhile $\mathbb{Z}_2$ symmetry that would prevent trilinear terms in the superpotential and which is only broken by loops of GUT scale particles, then we might expect $\lambda \sim \mathcal{O}(10^{-4})$, much as in the case described in Sec.~\ref{hiddenphoton}. Alternatively, one may take the view that this coupling is dimensionless and thus only logarithmically renormalized; therefore, small values are acceptable from the effective field theory perspective. Regardless of the justification, we find that the couplings must be small in order to effectively hide the NMSSM singlet sector from LUX and other direct detection constraints.

Although mono-jet, mono-$b$ and other commonly studied dark matter search channels at the LHC are highly suppressed in this model (and in the hidden photon model), decays of the Higgs into singlino pairs could provide a potentially observable signal. We find, however, that present constraints on the invisible width of the Higgs \cite{Campbell:2013una,Khachatryan:2014iha} are less sensitive than those imposed by direct detection searches by at least an order of magnitude.

\section{Discussion and Conclusions}
\label{conclusion}

The Galactic Center gamma-ray excess poses an intriguing set of challenges to contemporary particle physicists. Of particular interest are the following questions: what particle dark matter models are capable of producing the observed gamma-ray signal while also evading constraints from direct detection experiments, and what are the observational consequences that can be used to distinguish between these models?

The possibility explored in this paper is that the dark matter is part of a hidden sector which is imperfectly secluded from the Standard Model. If there is a hidden sector force, then the gauge boson that communicates that force may kinetically mix with the photon of electromagnetism, thereby attaining small couplings to those Standard Model fields that carry electric charge. Alternately, if the hidden sector masses are generated by a new Higgs field, then the hidden sector Higgs gauge eigenstate may undergo mass mixing with the Standard Model Higgs, and thus could communicate to the Standard Model via Yukawa couplings. Regardless of how the mixing occurs, dark matter annihilation in these models proceeds in two steps: first, two dark matter particles annihilate into on-shell intermediate hidden sector states, followed by the decay of those states into Standard Model particles. This two step annihilation setup makes it possible for the dark matter to annihilate at the rate required to produce the observed gamma-ray excess, while possessing almost arbitrarily small couplings to the Standard Model.

In this paper, we have explored two distinct theoretical settings that can accommodate this kind of model building. Within the context of a hidden sector endowed with a new abelian force, we can fit the gamma-ray excess when the hidden gauge boson kinetically mixes with the Standard Model photon. This model remains compatible with direct detection constraints as long as this kinetic mixing is small, $\epsilon \lsim  {\cal O} (10^{-4})$. The range of kinetic mixing anticipated to be induced by one-loop processes will be probed by operating and upcoming direct detection experiments, such as LUX and XENON1T.  We have also considered the gamma-ray excess within the context of the generalized NMSSM. By fixing the coupling $\lambda$ to values of $\sim$${\cal O}(10^{-3})$ or less, we can sufficiently sequester the Higgs singlet and its superpartner (a singlino-like neutralino) to evade direct detection constraints, while still generating the observed gamma-ray excess. Similar to the kinetic mixing scenario, it is plausible that next generation direct detection experiments will be sensitive to this class of models.

\bigskip \bigskip \bigskip

{\it Acknowledgements}: We would like to thank Matt Buckley, Jong-Chul Park, Tracy Slatyer and Kathryn Zurek for helpful discussions. DH is supported by the Department of Energy. AB is supported by the Kavli Institute for Cosmological Physics at the University of Chicago through grant NSF PHY-1125897. PG is supported by the National Research Fund Luxembourg through grant BFR08-024. SDM is supported by the Fermilab Fellowship in Theoretical Physics. Fermilab is operated by Fermi Research Alliance, LLC, under Contract No.~DE-AC02-07CH11359 with the US Department of Energy.

\bibliography{hiddengev}

\end{document}